\documentclass[3p,12pt]{revtex4}
\usepackage{bm}
\usepackage{amsfonts}
\usepackage{amssymb}
\usepackage{amsmath}
\usepackage{pdfsync}
\usepackage{graphicx}

\newcommand{\greeksym}[1]{{\usefont{U}{psy}{m}{n}#1}}
\newcommand{{\rmssmu}}{\mbox{\scriptsize{\greeksym{m}}}}

\newcommand{\rd}{{\rm d}}

\newcommand{\iO}{{\it \Omega}}

\fontfamily{lmss}\selectfont
\begin{document}

\title{Reply to the Comment on ``Magnetic moments in the Poynting theorem, Maxwell
equations, Dirac equation, and QED''}

\newcommand{\addrGaithersburg}{National Institute of Standards and Technology,
Gaithersburg, MD 20899-8420, USA}

\author{Peter J.~Mohr}
\affiliation{\addrGaithersburg}

%\date{\today}

%\keywords{Poynting theorem, Maxwell equations, Dirac equation, QED,
%magnetic moment}

\begin{abstract}

This reply addresses concerns expressed in a comment
(arXiv:2508.16689v1) on the paper given in the title (arXiv:2501.02022v2),
hereafter referred to as the paper.

%{\tt files in Documents/Papers/QEM/MagDipole}

\end{abstract}
%
%\pacs{03.50.De, 92.60.Ta, 11.30.Cp}
%
\maketitle
%
%%%%%%%%%%%%%%%%%%%%%%%%%%%%%%%%%%%%%%%%%%%%%%%%%%%%%%%%%%%%%%%%%%%%%

%\tableofcontents

The comment asks why a particular angular average was given for the
magnetic dipole field.  The reason is that it is the appropriate one for
the longitudinal magnetic dipole field.  There are two expressions for
the field.  One is transverse $\bm B_{\bm m}^{\rm T}(\bm x)$ and another
is longitudinal $\bm B_{\bm m}^{\rm L}(\bm x)$, where
\begin{eqnarray}
\bm\nabla\cdot\bm B_{\bm m}^{\rm T}(\bm x) = 0,
\qquad
\bm\nabla\times\bm B_{\bm m}^{\rm L}(\bm x) = 0.
\end{eqnarray}
For $|\bm x|>0$ they are equal:
\begin{eqnarray}
\bm B_{\bm m}^{\rm T}(\bm x) = \bm B_{\bm m}^{\rm L}(\bm x)
=
\frac{\mu_0}{ 4 \pi}  \,
\frac{3 \bm{\hat x}(\bm{\hat x}\cdot \bm m) - \bm m }
{ |\bm x|^3} \qquad\mbox{for}\quad |\bm x|>0.
\end{eqnarray}
However, they have different behavior at $\bm x = 0$.  This is
shown by taking the angular average of each expression.  For $|\bm
x|>0$ the angular average vanishes for both.  On the other hand, the
angular average reveals the behavior at $\bm x = 0$
\begin{eqnarray}
\frac{1}{4\pi}\int\rd\iO\,\bm B_{\bm m}^{\rm T}(\bm x)
=  \frac{2\mu_0}{3}\,\bm m\,\delta(\bm x) ,
\qquad
\frac{1}{4\pi}\int\rd\iO\,\bm B_{\bm m}^{\rm L}(\bm x)
=  - \frac{\mu_0}{3}\,\bm m\,\delta(\bm x) .
\end{eqnarray}
Derivations of both of these expressions appear in the paper and will
not be repeated here.  Also Eq.~(144) of the paper gives
\begin{eqnarray}
\bm B_{\bm m}^{\rm T}(\bm x) &=&
\bm B_{\bm m}^{\rm L}(\bm x) + \mu_0\,\bm m \, \delta(\bm x) .
\end{eqnarray}

The subsequent statement in the comment:``This means that Eq. (17) of
PJM ...  is wrong because it must be replaced by the angular average''
contains two errors.  First, Eq.~(17) of the paper is correct and
second, there is no reason why it should be replaced by its angular
average.  The conclusion that the rest of the paper is wrong based on
the remarks in the comment is incorrect.  The confusion in the comment
appears to be based on the assumption that all magnetic dipole fields
are transverse, which is not true.  Incidentially, Eqs.~(3), (4), and
(6) of the comment are missing factors of $1/(4\pi)$.

\end{document}